\documentclass[12pt]{iopart}
\usepackage{iopams}
\usepackage{bm}
\usepackage{mathbbol}
\usepackage[dvips]{graphicx}
\usepackage{color}
\usepackage{ulem}

\newcommand{\tbtio}{Tb$_2$Ti$_2$O$_7$ }

\newcommand{\tbi}{Tb$^{3+}$ }
\newcommand{\bo}{$\bar{1}$}

\newcommand{\ti}{${\it T}$}

\begin{document}


\title{Proposal for a [111] Magnetization Plateau in 
the Spin Liquid State of Tb$_2$Ti$_2$O$_7$}

\author{H R Molavian$^{1}$, and M J P Gingras$^{1,2}$}

\address{$^{1}$ Department of Physics and Astronomy, University of Waterloo,
 Waterloo, Ontario, N2L 3G1, Canada. }

\address{$^{2}$ Canadian Institute for Advanced Research, Toronto, Ontario, M5G 1Z8, Canada. }

\ead{gingras@gandalf.uwaterloo.ca}

\begin{abstract}
Despite a Curie-Weiss temperature $\theta_{\rm CW} \sim -14$ K, the
Tb$_2$Ti$_2$O$_7$ pyrochlore magnetic material lacks
long range magnetic order down to at least $T^*\approx 50$ mK.
It has recently been proposed that the low temperature collective paramagnetic
or spin liquid regime of this material may be akin to a spin ice state
subject to both thermal and quantum fluctuations $-$ a {\it quantum spin ice}
(QSI) of sorts.
Here we explore the effect of a magnetic field ${\bm B}$ along the $[111]$
direction on the QSI state.
To do so, we investigate the magnetic properties of a microscopic
model of Tb$_2$Ti$_2$O$_7$ in an independent tetrahedron approximation in a 
finite ${\bm B}$ along $[111]$. 
Such a model describes semi-quantitatively the collective paramagnetic regime where nontrivial
spin correlations start to develop at 
the shortest lengthscale, that is over a single tetrahedron, but where no long range order is
yet present.
Our results show that a magnetization plateau develops at low temperatures as the
system develops ${\bm B}=0$ ferromagnetic spin-ice-like ``two-in/two-out'' correlations at 
the shortest lengthscale. From these results, we are led to propose that
the observation of such a [111] magnetization plateau in 
Tb$_2$Ti$_2$O$_7$ would provide compelling
evidence for a QSI at ${\bm B}=0$ in this material and 
help guide the development of a theory for
the origin of its spin liquid state. 


\end{abstract}


\section{Introduction}

The spin liquid (SL) is the state of a 
magnetic system
characterized by short range spin-spin correlations 
down to temperatures much lower than the Curie-Weiss temperature, $\theta_{\rm CW}$, 
which is set by the magnetic interactions. 
While the SL concept was originally proposed
in 1973 in the context of a 
 spin-1/2 Heisenberg antiferromagnet on a triangular lattice,~\cite{Anderson}
it is only in recent years that 
its systematic experimental search in highly
frustrated magnetic systems has really taken off.~\cite{PhysicsToday}
A number of quasi 
two-dimensional~\cite{2DSL}
and three-dimensional ($3D)$ \cite{3DSL,Gardner-PRL}
materials with SL behavior
have been tentatively identified.
Yet,  the \tbtio insulator is 
perhaps one of the $3D$  highly frustrated magnetic systems
 with SL behavior that has been attracting interest for the longest time.~\cite{Gardner-PRL}
In  Tb$_2$Ti$_2$O$_7$, the magnetic Tb$^{3+}$ ions reside
on a pyrochlore lattice of corner-sharing tetrahedra.
Despite $\theta_{\rm CW}\approx -14$ K,~\cite{Gingras-PRB-2000}
Tb$_2$Ti$_2$O$_7$ does not develop conventional magnetic long range order
down to at least $T^*=50$ mK.~\cite{Gardner-PRL,Gardner-PRB,Gardner-PRB-50mK,Yasui}

It is unclear why Tb$_2$Ti$_2$O$_7$ fails to order.~\cite{Reviews}
It was originally thought that the Tb$^{3+}$ magnetic moments could be
described by classical Ising spins that can only
point in or out of a given tetrahedron,~\cite{Gingras-PRB-2000,Hertog-PRL}
similar to the Ho$^{3+}$ and Dy$^{3+}$  moments in
the Ho$_2$Ti$_2$O$_7$ \cite{Harris-PRL} and Dy$_2$Ti$_2$O$_7$ 
\cite{Ramirez-Nature}
spin ice materials.~\cite{Bramwell-Science}
However, such an Ising model for
 Tb$_2$Ti$_2$O$_7$  predicts a transition
to a four sublattice N\'eel ordered state at a temperature around 1 K,~\cite{Hertog-PRL}
in dramatic disagreement with experiments.~\cite{Gardner-PRL,Gardner-PRB-50mK}
The microscopic justification for depicting  the Tb$^{3+}$ 
moments as Ising spins stems from considering only
the single ion crystal field (CF) 
ground state doublet of Tb$^{3+}$ and neglecting  higher excited levels  
when describing the system at temperatures $T\lesssim 10$ K.~\cite{Gingras-PRB-2000}
However, a number of experimental~\cite{Gardner-PRB,Yasui,Rule} and 
theoretical~\cite{Kao,Curnoe} results argue against such an Ising description.
For example, an Ising model fails to explain, even at a qualitative level,
the symmetry of the neutron scattering intensity pattern, $I({\bm q})$, 
even in the paramagnetic phase.~\cite{Gardner-PRB,Kao}
Interestingly, $I({\bm q})$ remains largely unaltered from 4 K ~\cite{Gardner-PRB}
down to 50 mK, never extending much beyond the size of a tetrahedron primitive
basis.~\cite{Gardner-PRB-50mK}
One must therefore 
incorporate the excited 
CF states
when constructing an effective low-energy Hamiltonian, $H_{\rm eff}$, describing
Tb$_2$Ti$_2$O$_7$ at $T\lesssim 10$ K.
The main reason for this necessity is that,
unlike in spin ices,~\cite{Rosenkranz} the exchange and dipolar interactions
in Tb$_2$Ti$_2$O$_7$
are not much smaller than the energy gap $\Delta\approx 18$ K
separating the ground and first excited CF 
doublets.~\cite{Gingras-PRB-2000,Gardner-PRB,Mirebeau-INS} 
Also, when described as a classical 
$\langle 111\rangle$ Ising model, Tb$_2$Ti$_2$O$_7$ is found to be
close to the boundary between an ``all-in/all-out''
four sublattice N\'eel ordered ground state and a 
spin ice state.~\cite{Gingras-PRB-2000,Hertog-PRL}
Hence, perturbations, such as those
introduced by virtual excitations to the excited crystal field states,~\cite{Molavian}
can in principle give rise to new ground states,
and these excitations must therefore be considered at the very outset.

A recent work \cite{Molavian} argued,
on the basis of  perturbation theory 
calculations that consider an independent
tetrahedra approximation (ITA) to model the SL state of Tb$_2$Ti$_2$O$_7$, that
exchange and magnetic dipole-dipole interactions induce significant admixing
between the ground and excited CF  states.
Most importantly, this admixing
renormalizes the Ising (longitudinal) part of $H_{\rm eff}$
from that of a non-frustrated $\langle 111 \rangle$
Ising antiferromagnet~\cite{frust-FM} to that
of a frustrated  $\langle 111 \rangle$ Ising
ferromagnet, in essence turning the  system into a spin
ice.~\cite{Harris-PRL,Bramwell-Science,frust-FM,Molavian_unpub,entangled}
Interaction-induced admixing of the CF states
also generates 
effective off-diagonal (transverse) 
spin-spin interactions in $H_{\rm eff}$.~\cite{Molavian,Molavian_unpub}
As a result, \tbtio may possibly be described
by an effective frustrated $ \langle 111 \rangle$ Ising ferromagnet model with additional
quantum fluctuations transverse to the local 
$\langle 111 \rangle $ directions.~\cite{Molavian_unpub}
In other words, Tb$_2$Ti$_2$O$_7$ may be viewed in the temperature range 
[50 mK $\lesssim T \lesssim$ 10 K]
as being in a thermally and 
quantum mechanically fluctuating spin ice state where
frustrated ferromagnetic correlations in the $\langle 111\rangle$ Ising components
exist at the shortest lengthscale $-$ 
a {\it quantum spin ice} (QSI) of sorts.~\cite{QSI-explain}
Hence, Tb$_2$Ti$_2$O$_7$ may constitute 
an exciting playground to explore quantum  effects in a spin ice system with
the QSI proposal providing a long awaited and useful theoretical perspective 
to rationalize Tb$_2$Ti$_2$O$_7$.~\cite{Reviews}
The possibility of a QSI in Tb$_2$Ti$_2$O$_7$ is also of somewhat fundamental
interest as it may relate to the topical problem
of fractionalized spin excitations and emerging photon-like excitations
in theoretical models of frustrated quantum Ising antiferromagnets
on the pyrochlore lattice. \cite{S12pyro}

%
%

Because of the accidental quasi-cancellation of the exchange
and dipolar energy scales at the nearest-neighbor level, 
the zero-frequency QSI correlations 
in Tb$_2$Ti$_2$O$_7$ may be masked 
by strong dynamical fluctuations and difficult to
detect via neutron scattering~\cite{Gardner-PRL,Gardner-PRB,Gardner-PRB-50mK,Yasui} 
and muon spin relaxation experiments.~\cite{Gardner-PRL}
From this perspective, the behaviors exhibited by 
Tb$_2$Ti$_2$O$_7$  may be viewed as an interesting
example of the spectral weight downshift
expected in highly frustrated magnetic systems.~\cite{Ramirez}
We believe that this phenomenology 
is an important part of the physics of Tb$_2$Ti$_2$O$_7$ and at the 
origin of the stumbling blocks in 
developping a microscopic theoretical
understanding of Tb$_2$Ti$_2$O$_7$.~\cite{Reviews}
With this in mind, we explore here whether low temperature 
bulk magnetization, ${\bm M}(T)$, measurements 
may be used to expose whether the nearest-neighbor part of 
the  longitudinal (Ising) sector of $H_{\rm eff}$ describing Tb$_2$Ti$_2$O$_7$  
is indeed that of a frustrated $\langle 111 \rangle$ ferromagnet~\cite{frust-FM} and,
 therefore,
whether a QSI description of this system is correct.
Perhaps one of the clearest tell-tale bulk magnetic signatures of
frustrated ferromagnetic ``two-in/two-out'' ice-rule 
correlations~\cite{Bramwell-Science} in  a
spin ice material is the emergence of a magnetization plateau for a magnetic
field ${\bm B}$ along the
$[111]$ direction (diagonal of the conventional cubic unit cell)
at low temperatures once the ice-rule obeying spin correlations
have  become established.~\cite{Sakakibara-PRL,Moessner-111}
Motivated by this magnetization plateau phenomenon in classical Ising
spin ices, 
we therefore ask here whether a plateau-like feature in the
$[111]$ magnetization can also develop in the collective 
paramagnetic regime of Tb$_2$Ti$_2$O$_7$ as the frustrated
ferromagnetic $\langle 111\rangle$ spin correlations 
develop at the shortest (single tetrahedron) lengthscale.

The suggestion that Tb$_2$Ti$_2$O$_7$ may be described at low energies
and low temperatures by an effective Hamiltonian, $H_{\rm eff}$, whose 
longitudinal Ising components are frustrated by effective nearest-neighbor
couplings was motivated by perturbation theory calculations carried out over
a single tetrahedron of interacting Tb$^{3+}$ ions.~\cite{Molavian}
The coupling between the $\langle 111 \rangle$ Ising components for a pair
of interacting Tb$^{3+}$ ions is in principle renormalized by  the interactions
with all the other Tb$^{3+}$ ions in the system.
Therefore, an important question is whether the frustrated
``two-in/two-out'' spin-ice-like correlations are preserved in a perturbation theory
extended to a thermodynamically large lattice and, in particular, in the presence
of the long-range magnetostatic dipole-dipole interactions between Tb$^{3+}$ 
ions.~\cite{Gingras-PRB-2000,Hertog-PRL,Kao}
Our preliminary 
perturbation theory calculations carried over the whole lattice show that the 
system remains in an ice-rule obeying ground state with competing states with
a ``three-in/one-out'' spin configuration having a slightly higher 
energy.~\cite{Molavian,Molavian_unpub}
Consequently, we would expect that, in an ``exact'' treatment of $H_{\rm eff}$ at 
nonzero temperature, spin ice ``two-in/two-out'' correlations would develop
at low temperatures. The important
question is therefore whether such correlations
actually develop in real Tb$_2$Ti$_2$O$_7$ at low temperatures
and if 
they can be evinced by the observation of a $[111]$ magnetization plateau.

To get a preliminary handle on the relevant temperature
and magnetic field scale where QSI physics may become experimentally 
discernable in  Tb$_2$Ti$_2$O$_7$, we consider the same ITA
used in Ref.~\cite{Molavian} to describe the SL 
state of that material since, as stated above,
interactions beyond a single tetrahedron do not spoil
the existence of an ice-rule obeying semi-classical ground state.~\cite{Molavian_unpub}
The merit of the ITA to describe the collective paramagnetic state
of Tb$_2$Ti$_2$O$_7$ was discussed in Refs.~\cite{Molavian,Molavian_unpub,Curnoe-ITA}.
In the context of the work presented here, one could interpret the ITA as a self-consistent
cluster mean-field theory (SCMFT)~\cite{SCMFT} where the considered cluster is
a tetrahedron, {\it but} where the self-consistency conditions that incorporate the inter-clusters
(inter-tetrahedra) interactions have been set to zero. This is because we are describing the
collective paramagnetic/spin liquid regime of the model 
where the correlations extend solely over the smallest cluster lengthscale where 
no global symmetry is (yet) spontaneously broken.
The experimental observation that spin correlations in Tb$_2$Ti$_2$O$_7$ never develop
beyond the size of a single tetrahedron from a temperature of $\sim 10^0$ K down to 50 mK 
~\cite{Gardner-PRL,Gardner-PRB,Gardner-PRB-50mK,Yasui}
would  seem to further provide a post-factum justification to this ``constrained'' paramagnetic
description of a SCMFT of the
$H_{\rm eff}$ of Tb$_2$Ti$_2$O$_7$.~\cite{PM-justification}
In other words, the ITA should be qualitatively valid
for sufficiently small field and temperatures such that long-range 
spin-spin correlations are not induced.~\cite{Rule}
Finally, for the proposed experiment seeking to expose a magnetization plateau for
${\bm B}$ along [111], symmetry considerations play a crucial 
role,~\cite{Sakakibara-PRL,Moessner-111}
and the ITA does capture that symmetry aspect of the problem.
Below, we find that for parameter values for which the model exhibits a QSI,
and which are appropriate for Tb$_2$Ti$_2$O$_7$,~\cite{Molavian} 
the low-temperature magnetization $M(T)$ 
($T\lesssim 100$ mK) does indeed exhibit a 
plateau for  $B \equiv \vert {\bm B} \vert  \sim 0.1$ T.


\section{Model and Results}

Within the ITA, the Hamiltonian for Tb$_2$Ti$_2$O$_7$ reads

\begin{eqnarray}
H&=& \sum_{{a=1},{b > a}}^4
(H_{ab}^{\rm ex}+H_{ab}^{\rm dip})
+\sum_{a=1}^4 ( H_a^{\rm Z} + H_a^{\rm cf} )  .
\label{hamil}
\end{eqnarray}
$H_{ab}^{\rm dip}={\cal D}R_{\rm nn}^3 
\left[ {{\rm\bf J}_a} \cdot {{\rm\bf J}_b}
 - 3({\rm\bf J}_a \cdot {\hat r}_{ab})
({\rm\bf J}_b \cdot {\hat r}_{ab})
 \right]{|{\bm R}_{ab}|^{-3}}$
and 
$H_{ab}^{\rm ex}={\cal J}{\bf J}_a\cdot{\bf J}_b$
describe, respectively, the dipole-dipole and the exchange interactions
for a pair $ab$. $H_a^{\rm Z}=-g\mu_{\rm B}{\bf J}_a\cdot{\bm B}$ 
is the Zeeman Hamiltonian  and
$H_a^{\rm cf}$ is the crystal field (CF)
Hamiltonian, both for ion ${\it a}$.~\cite{Gingras-PRB-2000,Mirebeau-INS} 
${\bf R}_{ab} \equiv {\bf R}_b-{\bf R}_a=|{\bf R}_{ab}|{\hat r}_{ab}$ 
where ${\bf R}_a$ is the position 
vector of ion $a$, and $R_{\rm nn}=3.59 \AA$ is the nearest-neighbor distance. 
${\cal D}=\mu_0(g \mu_{\rm B})^2/(4\pi {R_{\rm nn}}^3) =0.0315$ K is the 
dipole-dipole coupling and 
${\cal J}=0.167$ K is the exchange coupling, with the convention that 
${\cal J} > 0$ is antiferromagnetic.~\cite{Molavian,theta-critique}
$\mu_{\rm B}$ is the Bohr magneton, 
$g=\frac{3}{2}$ is the Land\'e factor of Tb$^{3+}$ and
${\bf J}_a$ ($|{\bf J}_a|=6$)
is the total angular momentum operator of Tb$^{3+}$ ion $a$.
Calculations considering a single  (non-interacting) 
Tb$^{3+}$ ion in nonzero ${\bm B}$ 
show that the six lowest energy CF states
describe well the exact ${\bm M}$ when compared with calculations
that consider all 2J+1=13 CF states 
for $B\lesssim 30$ T with an error less than 1\%. 
Henceforth, we use the six lowest CF states of \tbi to calculate ${\bm M}$ 
within the ITA. 
The basis states of the system are taken as the tensor product of 
the CF states of the four non-interacting \tbi ions on a tetrahedron. 
We denote the two states 
of the CF
ground state by
 $|\!\!\uparrow \rangle$ and 
$|\!\!\downarrow \rangle$. 
These  correspond, respectively, to the ``out'' and ``in'' 
degenerate (Ising) CF ground states of
$H^{\rm cf} \equiv \sum_a H_a^{\rm cf}$ 
on an ``upward pointing'' tetrahedron 
primitive basis~\cite{Gingras-PRB-2000,Hertog-PRL,Sakakibara-PRL}
 when referring to the $[111]$ direction along which ${\bm B}$ points.
The CF ground doublet and first excited doublet
 are given in Ref.~\cite{Molavian} 
and the two other excited states,
determined as in Ref.~\cite{Molavian},
 are
$|\psi^{(3)} \rangle=
 a_6 (| 6 \rangle + |-6 \rangle)+
 a_3 (| 3 \rangle - |-3 \rangle)$ 
and 
$|\psi^{(4)} \rangle=
  b_6(| 6 \rangle + |-6 \rangle) +
  b_3(| 3 \rangle - |-3 \rangle) + 
  b_0 | 0 \rangle$
with  energies $E_3=142.1$ K and $E_4=212.4$ K, respectively.~\cite{coeff}
Here we express 
$|\psi^{(3)} \rangle$ and
$|\psi^{(4)} \rangle$ 
in terms of the eigenstates
$\vert  {\rm J} =6, m_{\rm J}\rangle$
of ${\rm J}^z$ within the fixed ${\rm J}=6$ manifold.
The local $\hat z$ easy axes for sublattices  
1 to 4 on an individual tetrahedron 
primitive unit cell
are along
 [111], [\bo\bo1], [1\bo\bo] and [\bo1\bo] and we 
maintain this sequence of labelled spins/sites throughout the paper, with
$|ijkl\rangle \equiv |ijkl\rangle_{1234} \equiv
|i\rangle_1\otimes|j\rangle_2\otimes|k\rangle_3\otimes|l\rangle_4$.
  
We determine the matrix elements of the full $H$ in Eq. (1) 
in the CF basis, find 
the eigenvalues and eigenstates of $H$ using a numerically   
exact diagonalization method and use these
to calculate ${\bm M}$ 
given by the standard formula
\begin{eqnarray}
M^{\alpha}&=& \frac{g \mu_{\rm B}}{4} \sum_{i=1}^{N_s} \sum_{a=1}^4 u_a^{\alpha \beta} 
\langle \phi_i |  {\rm J}_a^{\beta} | \phi_i \rangle n_{\rm B}(E_i)  \; ,
\end{eqnarray}
where $M^{\alpha}$ is the $\alpha$ component of the total ${\bm M}$ per ion.
 $|\phi_i \rangle$'s are
the eigenstates of Hamiltonian (\ref{hamil}), $u_a^{\alpha\beta}$ is 
the $\alpha\beta$ component 
of the rotation matrix from the local quantization frame,
aligned along the local cubic [111] direction at site $a$, 
to the global axis.
 ${{\rm J}}_a^\alpha$ is the $\alpha$ component 
of ${\rm\bf J}$
expressed in the local $x y z$
quantization frame with ${\hat z}$ aligned with
the local $[111]$ axis.
$n_{\rm B}(E_i)$=$\exp({-E_i}/{k_{\rm B}T})$ is the
Boltzmann population of state ${\it i}$ with energy $E_i$ at temperature $T$. 
The sum over $i$ in Eq. (2) is carried over
the $N_s=6^4=1296$ eigenstates retained in $H^{\rm cf}$
and the sum on ${\it a}$ is over the four sites of a tetrahedron.

\begin{figure}[htbp]
\centerline{\scalebox{0.4}{\includegraphics{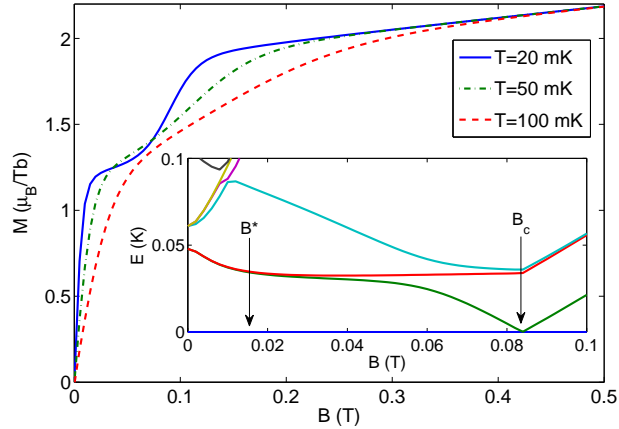}}}
\caption{(Color online).  Magnetization ${\bf M}$ of \tbtio in the ITA as a 
function of magnetic field ${\bm B}$ applied along the $[111]$ direction. 
In the magnetization 
plateau regime, where $M(B)$ flattens out 
for $T=20$ mK, the  system is in a ``quantum kagome ice'' state
(see text). Inset: 
energy of the lowest states as a function of $B$ which 
which shows a level-crossing at $B_c$=0.082 T.}
\label{kagome}
\end{figure}

We calculate ${\bm M}$ within the ITA for ${\bm B}$ along
$[111]$ and for $T$=20, 50 and 100 mK.
The result of this calculation is shown in Fig.~\ref{kagome}.
As in spin ices,~\cite{Sakakibara-PRL}
a magnetization plateau occurs, becoming apparent
only below the temperature at which the ice rules
are well established.~\cite{Molavian}
That is, when the temperature drops below the very small energy gap
separating the manifold of ``two-in/two-out'' states from
the manifold of the ``three-in/one-out'' states.
In the present model,  the spin-ice phenomenology is ``hidden''
at low temperature because of the accidental location of
the system near the phase boundary between the
``all-in/all-out''  (doublet) N\'eel  
state  and the ``two-in/two-out''spin ice state.~\cite{Molavian}
The plateau develops only at low temperature because
the QSI state in $H$ in Eq. (1), and possibly
in Tb$_2$Ti$_2$O$_7$, is inherited by perturbative 
virtual quantum mechanical 
crystal field excitations renormalizing the already 
accidentally almost vanishing classical (longitudinal Ising) 
interactions in
$H_{\rm eff}$.~\cite{Molavian} 
Hence, the manifold of 
low-energy states at ${\bm B}=0$, which
descend from the
$2^4=16$ parent Ising CF ground doublet states, span a small energy
bandwidth $\delta \Omega \sim  0.5$ K,~\cite{Molavian} with the 6 
lowest energy  spin ice states 
spanning an even smaller bandwidth $\delta \omega \sim 0.06$ K
(see the six lowest 
[singlet $\oplus$ doublet $\oplus$ triplet] states at ${\bm B}=0$ in
the inset of Fig. 1).~\cite{Molavian}
Consequently, the plateau in $M$  only
emerges at quite low temperature. This constitutes a prime example
where high frustration is caused by naturally 
accidentally fine-tuned
microscopic interactions (here
$H^{\rm ex}$ and $H^{\rm dip}$)
and where the relevant low-energy physics is pushed
down to an exceedingly small scale.~\cite{Ramirez}
For small $B$, $M$ starts 
to increase even at a very low $T$
(e.g. 20 mK).
The reason for this is that, despite the ground state being a singlet 
at the ITA level,  there exist a second order (van Vleck like)
susceptibility originating from the excited states which leads
to $M\ne 0$ for $B>0$.

\begin{figure}[htbp]
\centerline{\scalebox{0.4}{\includegraphics{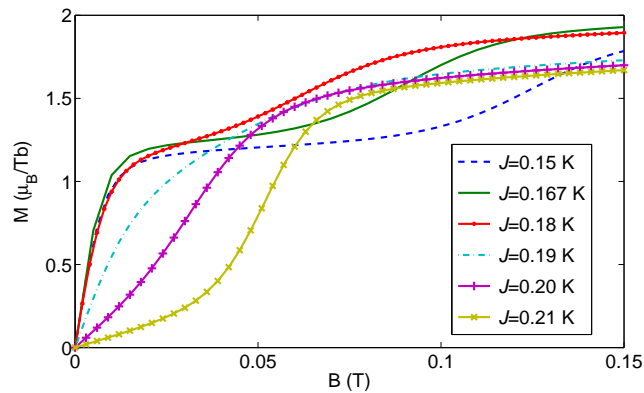}}}
\caption{(Color online).
[111] magnetization $M$ of \tbtio in the independent tetrahedron approximation as a 
function of $B$ 
in the $[111]$ direction.
Here the temperature is 20 mK. 	}
\label{kagomeJ}
\end{figure}

The value of the exchange ${\cal J}$ in \tbtio is not
fully agreed upon.~\cite{Gingras-PRB-2000,theta-critique}
It is therefore worthwhile to explore how $M(B)$ 
changes as ${\cal J}$ is varied.
We plot in 
Fig.~\ref{kagomeJ} $M$ 
as a function of ${B}$ 
for \ti= 20 mK and exchange 
couplings ${\cal J}$ = 0.15, 0.167, 0.18, 0.19, 0.20 K.
These results show that a  magnetization 
plateau is manifest only
when the system displays a QSI state
at ${\bm B}=0$ for ${\cal J} < {\cal J}_{\rm c}$.~\cite{Molavian}
Here ${\cal J}_{\rm c}$=0.187 K 
delineates the transition between the 
QSI  and the ``all-in/all-out'' (doublet) state,~\cite{Molavian}
 with the plateau disappearing
for ${\cal J} > {\cal J}_{\rm c}$.
  For ${\cal J}=0.18$ K, the system is barely 
beyond the threshold for exhibiting a QSI
 state and one finds 
that the plateau is poorly formed even at $T=20$ mK.
Our calculations also show that, although a rescaling of ${\cal J}$
and ${\cal D}$ changes  the magnetization plateau value and the
$B$ range over which a magnetization plateau arises, 
it does not change the 
generic qualitative behavior seen in Fig. 1 as
long as the model, and presumably the real system, 
displays a QSI state for ${\bm B}=0$ or, in other words, 
as long as the system possesses 
frustrated ferromagnetic interactions in the nearest-neighbor
Ising part of $H_{\rm eff}$.

\section{Field Dependence of Energy Levels on a Single Tetrahedron}

The ITA results above can be interpreted as describing
the behavior of the system in the temperature range
where no long
range order exists except for the field-driven
global $[111]$ magnetization, but 
where spin correlations extend over the shortest 
lengthscale characteristic of a collective paramagnetic/spin liquid regime 
(i.e. over a tetrahedron). Yet, it is interesting, for completeness, to
consider the evolution of the energy levels over a single tetrahedron
as this evolution would ultimately need to be taken into account in
a SCMFT ~\cite{SCMFT}
 going beyond a single tetrahedron to describe the possible spontaneously 
broken symmetry states of the model.

To understand  $M(B)$  at 20 mK within the ITA,
 we scrutinize the ground state of the system as a function of ${ B}$. 
We first note that the ground state, 
$|\Psi\rangle$,
in the $B$ range considered in Fig.~\ref{kagome},
can be generically and compactly written as
$|\Psi\rangle=\alpha|\omega\rangle+\bar \alpha|\bar \omega\rangle+
\beta |\!\!\uparrow \downarrow \downarrow \downarrow \rangle
+\epsilon |\chi \rangle$.
Here 
$|\omega\rangle=|\!\uparrow \uparrow \downarrow \downarrow \rangle+
|\!\uparrow \downarrow \uparrow \downarrow \rangle
+|\!\uparrow \downarrow \downarrow \uparrow \rangle$
and $|\bar \omega\rangle$ is the time conjugate of $|\omega\rangle$.  
$\alpha$, $\bar \alpha$, $\beta$, $\epsilon$ and 
$|\chi \rangle$ are $B$-dependent and $\epsilon$ is 
much smaller than at least one of the coefficients 
$\alpha$, $\bar \alpha$, or $\beta$.  
Notice that in $|\omega\rangle$, ion \#1 is in the 
${|\!\uparrow\rangle}$  (``out'') state. 
We denote the QSI state with ${ \bm B}=0$ in Eq. (1) as
$\vert {\rm QSI}\rangle \equiv \vert \Psi \rangle$ above with 
$\alpha= \bar \alpha$, $\beta =0$ and $| \epsilon |\ll | \alpha |$.

The evolution of the ground state with ${B}$ can be explained
from the competition between $H^{\rm Z}$ and
$H^{\rm ex}+H^{\rm dip}$ in Eq.~(1).
For weak ${B}$, 
$\langle {\rm QSI}| H^{\rm Z} | \varphi \rangle \neq 0$, where
$|\varphi \rangle$ refers to
any of the six pure  ``two-in/two-out''  states (e.g.
\mbox{$|\!\uparrow \uparrow \downarrow \downarrow \rangle$},
\mbox{$|\!\uparrow \downarrow \uparrow \downarrow \rangle$},
etc).
Among these, the three states with ion \# 1 in 
\mbox{$\! \vert \uparrow\rangle _1$}
(i.e. polarized along
[111])  have a lower Zeeman energy than 
the other three two-in/two-out states.  Hence,
with $B\ne 0$,
a state with larger weight from
these components, i.e. with ion \#1 \mbox{$\vert \uparrow\rangle _1$},
has lower energy than the 
${\bm B}=0$ $\vert {\rm QSI} \rangle$ state and become
the ground state.
As a result, for non-zero and sufficiently weak $B$,
the ground state evolves from $|{\rm QSI}\rangle$
to a state $|\Psi\rangle$  with $\alpha > \bar \alpha$ 
and with $\beta =0$ ($\beta =0$ because $H^{\rm Z}$ does not admix
$\vert {\rm QSI}\rangle$ with
\mbox{$|\!\uparrow \downarrow \downarrow \downarrow \rangle$} 
at any order in $H^{\rm Z}$).

As ${B}$ is increased, $\alpha$ grows
and $\bar \alpha$ decreases until
$B^* \approx 0.016$ T where 
 $\delta E^{\rm Z} \equiv 
\langle {\rm QSI}| H^{\rm Z} | {\rm QSI}\rangle \approx 0.06$ K. 
For $B\sim B^*$,
all 6 lowest energy 
states with a predominant two-in/two-out weight spanning
the small energy bandwidth $\delta \omega \approx 0.06$ K  
(see inset of Fig. 1) fall 
within the energy range $\delta E^Z$.
Hence, for $B>B^*$, the system adopts a ground state with
 $\alpha \gg \bar \alpha$, 
where ion \# 1 is polarized towards [111],
while the three other ions 
are in an entangled state which is a symmetric linear 
combination of two-in/one-out states.
This partially spin polarized (PSP) state, $\vert {\rm PSP}\rangle$,
is analogous to the kagome
ice state in classical spin ice systems~\cite{Sakakibara-PRL,Moessner-111}.
By increasing ${B}$ beyond $B^*$, up to $B_c$,
 the ground state remains
$\vert {\rm PSP} \rangle$ (i.e. $\beta=0$
 because $H^{\rm Z}$ does not admix 
$\vert {\rm PSP}\rangle$  with
\mbox{
\mbox{$ |\!\uparrow \downarrow \downarrow \downarrow \rangle$}}
at any order in $H^{\rm Z}$),
until a level crossing occurs for $B_{c} \approx 0.082$ T
(see inset of Fig.~\ref{kagome}).

One must consider the evolution 
of the first excited energy level with $B$ in order to understand
the above level crossing. 
For $B=0$, one of the two states from the first excited doublet has a small 
\mbox{$\vert \!\uparrow\downarrow\downarrow\downarrow\rangle$} 
``three-in/one-out'' component.
As a result, the matrix elements of $H^{\rm Z}$
between this 
state and the higher energy excited states which have 
a three-in/one-out component is nonzero. 
Hence, by increasing ${B}$, the 
\mbox{$\vert \!\uparrow\downarrow\downarrow\downarrow\rangle$} 
weight 
in the first excited state  
increases and its energy decreases and,
by the time $B \sim 0.08$ T, this state is almost pure
\mbox{$\vert \!\uparrow\downarrow\downarrow\downarrow\rangle$}.
This state then crosses the PSP 
state at $B_{c} \approx 0.082$ T and becomes the ground state
(see inset of Fig. (\ref{kagome}),
with the system exiting the PSP
state characterized by a magnetization plateau.

\section{Conclusion}
\label{Conclu}
    
We calculated the magnetization ${\bm M}$
of \tbtio for a field ${\bm B}$ along the [111] direction
within a single tetrahedron approximation (ITA). 
For weak ${B}$ in this direction, 
a magnetization plateau is found to occur 
for a temperature range where the system is in 
a collective paramagnetic/spin liquid regime. 
Within this regime, 
the correlations extend solely over the size of a tetrahedron and 
each tetrahedron primitive basis is in a state 
predominantly weighted by
\mbox{$|\!\!\uparrow\rangle_1 \otimes(|\!\!\uparrow\downarrow\downarrow \rangle_{234}+
|\!\!\downarrow\uparrow\downarrow\rangle_{234} + |\!\!\downarrow \downarrow 
\uparrow \rangle_{234})$},
where the three spins on the vertices of 
the triangle perpendicular to the [111] direction, i.e in the 
kagome plane, form
a linear combination of two-in/one-out states.~\cite{Sakakibara-PRL,Moessner-111} 
This partially spin-polarized state (PSP) is reminiscent of the 
classical kagome ice state in classical Ising 
spin ices~\cite{Sakakibara-PRL,Moessner-111}
which one may label {\it quantum kagome ice} to distinguish it from the  
degenerate kagome ice state.
From a microscopic level point of view, both the zero field 
quantum spin ice (QSI) and the PSP states originate 
from virtual crystal field excitations 
and quantum many-body effects.~\cite{Molavian}
At this time, low-temperature experiments are needed to test the
proposals of QSI and PSP states in Tb$_2$Ti$_2$O$_7$.
From the results presented here, we propose that
an experiment based on measurements of ${\bm M}$
along the [111] direction would help to ascertain the 
possibility of a QSI state in Tb$_2$Ti$_2$O$_2$.
Furthermore, the observation of a QSI and a PSP
in Tb$_2$Ti$_2$O$_7$ would confirm the status of this
material as a unique opportunity to explore quantum 
effects in a spin ice system.
It may also open an experimental avenue to explore some of the
exciting physics proposed to be at play in frustrated pyrochlore Ising
systems with perturbative quantum fluctuations.~\cite{S12pyro}
We hope that this work will motivate such investigations which would help
shed some light on the microscopic origin of the enigmatic 
spin liquid state in
Tb$_2$Ti$_2$O$_7$.~\cite{Gardner-PRL,Reviews} 

We thank M. Enjalran, P. Holdsworth, P. McClarty, R. Moessner 
and A. M. Tremblay for useful discussions.
This research was funded by NSERC and
the CRC program (Tier 1, MJPG).



\begin{thebibliography}{100}
\bibitem{entangled} The ground state of an isolated
single tetrahedron~\cite{Molavian}
is an entangled quantum variant of the 6-fold degenerate ``two-in/two-out'' 
classical ice-like spin configurations.~\cite{Harris-PRL,frust-FM}
\bibitem{PM-justification} That said, further theoretical work aimed to understand this
bottlenecked growth of spin correlations in Tb$_2$Ti$_2$O$_7$ below $\sim 10^0$ K is
most definitely warrented.
\bibitem{coeff}The values of coeffecients $a_m$ and $b_n$ are
 $a_6$=0.25, $a_3$=0.66, $b_6$=0.33, $b_3$=0.62 and $b_0$=0.12.
\end{thebibliography}
\end{document}